\documentclass[twocolumn,english,aps,prl,superscriptaddress,floats,nobibnotes,longbibliography]{revtex4-1}

\usepackage[utf8]{inputenc}
\usepackage{amsmath}
\usepackage{amssymb,bm}
\usepackage{amsthm}
\usepackage{setspace}
\usepackage{graphicx}
\usepackage{braket}
\usepackage{mathrsfs}
\usepackage{float}
\usepackage[colorlinks = true,linkcolor = blue,urlcolor  = blue,citecolor = blue,anchorcolor = blue]{hyperref}
\usepackage[english]{babel}
\usepackage{bm}
\usepackage{csquotes}
\usepackage{hyperref}

\begin{document}

\title{Chiral anomaly and planar Hall conductance in pseudospin-$1$ Fermions}
\author{Azaz Ahmad}
\affiliation{Department of Physics, Indian Institute of Technology Bombay, Powai, Mumbai 400076, India.}

\begin{abstract}
Positive longitudinal magnetoconductance (LMC) and planar Hall conductance (PHC) are widely regarded as hallmark transport signatures of the chiral anomaly in Weyl semimetals. Recent theoretical advances have generalized the concept of Weyl fermions to multifold Fermionic systems characterized by higher-pseudospin quasiparticle excitations, motivating a systematic investigation of their anomaly-induced transport responses. In this work, we employ semiclassical Boltzmann transport theory within the relaxation-time approximation to study magnetotransport in pseudospin-1 Weyl semimetals, incorporating momentum-dependent scattering, orbital magnetic moment corrections, and charge-conservation constraints. To study PHC, we explicitly break the azimuthal symmetry of the system through two mechanisms: (i) a generic tilt of the quasiparticle dispersion and (ii) a finite misalignment between the electric and magnetic fields. We find that, in the untilted case, the PHC is positive and exhibits a quadratic dependence on the magnetic field strength. Remarkably, increasing the scattering strength drives a sign reversal of the PHC, resulting in a transition from positive to negative values. Furthermore, the PHC displays the characteristic angular dependence $\sin 2\gamma$, where $\gamma$ denotes the angle between the applied magnetic field and the $x$-axis. The introduction of tilt qualitatively modifies this behavior: tilt along the $x$- and $z$-directions transforms the angular response into $\sin\gamma$ and $\cos\gamma$ forms, respectively, thereby inducing pronounced anisotropy in the planar Hall signal. In addition, the PHC exhibits a nonmonotonic dependence on the tilt magnitude, highlighting the intricate interplay between tilt-induced symmetry breaking and chiral-anomaly-driven transport in higher-spin Weyl systems. Our results provide experimentally accessible signatures of multifold fermions and establish a theoretical framework for interpreting magnetotransport measurements in candidate materials belonging to space groups 199, 214, and 220.
  
\end{abstract}

\maketitle
\section{Introduction}
\label{Introduction}
The concept of the chiral anomaly (CA) associated with Weyl fermions was originally identified as a key mechanism contributing to the anomalous decay processes of neutral pions~\cite{bertlmann2000anomalies}. The unambiguous experimental realization of Weyl fermions in condensed matter systems has, over the past decade, catalyzed extensive theoretical and experimental investigations~\cite{hosur2013recent,armitage2018weyl,Yan_2017,hasan2017discovery,burkov2018weyl,ong2021experimental,nagaosa2020transport,lv2021experimental,mandal2022chiral,ahmad2024geometry}. Foundational theoretical insights trace back to the seminal work of Nielsen and Ninomiya in the early 1980s, who rigorously demonstrated, within the framework of lattice gauge theory, that Weyl nodes must appear in pairs with opposite chirality to preserve both global charge conservation and gauge invariance~\cite{nielsen1981no,nielsen1983adler}. This phenomenon, commonly referred to as the Nielsen-Ninomiya theorem, underpins the topological stability of Weyl nodes in lattice systems. In the presence of external electromagnetic fields, however, the chiral current is not conserved—a non-perturbative quantum field theoretic effect known as the chiral anomaly, or more specifically, the Adler-Bell-Jackiw (ABJ) anomaly~\cite{adler1969axial}. In condensed matter systems, the chiral anomaly manifests as anomalous transport responses such as the negative longitudinal magnetoresistance and planar Hall effect, which serve as hallmarks of Weyl semimetals (WSMs). These signatures have been probed through diverse experimental modalities including magnetotransport, thermoelectric, and optical measurements~\cite{parameswaran2014probing,hosur2015tunable,goswami2015optical,goswami2013axionic,son2013chiral,burkov2011weyl,burkov2014anomalous,lundgren2014thermoelectric,sharma2016nernst,kim2014boltzmann,zyuzin2017magnetotransport,cortijo2016linear,das2019berry,kundu2020magnetotransport}. These studies not only substantiate the theoretical predictions surrounding the chiral anomaly but also reveal its profound implications for topological quantum materials and emergent low-energy excitations in solid-state systems.

While Weyl fermions originally emerged within the framework of high-energy physics, their realization in condensed matter systems has unveiled a broader class of Fermionic quasiparticle excitations that are not constrained by the Poincar\'{e} symmetry inherent to relativistic quantum field theories~\cite{bradlyn2016beyond,tang2017multiple,chang2018topological}. Owing to the crystalline symmetries present in solids, it becomes possible to host gapless Fermionic excitations beyond conventional Weyl fermions. The canonical low-energy effective Hamiltonian describing a Weyl Fermion is given by $H_\mathbf{k} \sim \chi \mathbf{k} \cdot \boldsymbol{\sigma}$, where $\boldsymbol{\sigma}$ represents the vector of Pauli matrices and $\chi = \pm 1$ denotes the chirality of the Fermion. In crystalline solids, however, one can encounter Fermionic excitations characterized by a generalized Hamiltonian of the form $H_\mathbf{k} \sim \chi \mathbf{k} \cdot \mathbf{S}$, where $\mathbf{S}$ denotes a higher-dimensional (pseudo)spin representation beyond spin-$\frac{1}{2}$, such as spin-1 or spin-$\frac{3}{2}$ matrices. These multifold degenerate chiral quasiparticles carry higher-order topological invariants, specifically nontrivial Chern numbers satisfying $|\mathcal{C}| > 1$, and act as monopole-like sources and sinks of the Berry curvature field in momentum space. This contrasts sharply with conventional Weyl fermions, which exhibit a twofold degeneracy with an associated Chern number of $|\mathcal{C}| = 1$. The existence of such higher (pseudo)spin fermions enriches the taxonomy of topological phases and expands the landscape of emergent quasiparticles in solid-state systems.

Although magnetotransport phenomena arising from the chiral anomaly (CA) and Berry curvature have been extensively explored in the context of Weyl fermions and their Kramers-Weyl counterparts~\cite{cheon2022chiral,das2023chiral,varma2024magnetotransport,ahmad2025longitudinal,varma2026chiral,varma2026strain,ahmad2026magnetotransport,ahmad2026semiclassical}, the systematic extension of these effects to quasiparticles with pseudospin higher than $1/2$ remains relatively underexplored. From a general theoretical standpoint, chiral quasiparticles with higher pseudospin are expected to exhibit distinct responses to external magnetic fields due to their enriched topological structure and higher-order Berry curvature contributions. Consequently, analogous magnetotransport signatures observed in pseudospin-$1/2$ systems are anticipated to persist in systems hosting higher-pseudospin excitations, albeit with quantitatively and qualitatively modified features.

For pseudospin-$1/2$ Weyl fermions, the application of external electromagnetic fields with a non-vanishing scalar product, $\mathbf{E} \cdot \mathbf{B} \neq 0$, results in the non-conservation of chiral charge, thereby inducing a chiral current through the chiral anomaly mechanism~\cite{son2013chiral}. Similar conclusions can also be drawn using semiclassical kinetic theory frameworks, incorporating Berry curvature corrections to the phase space dynamics. Motivated by these insights, recent theoretical efforts have focused on elucidating the transport characteristics of multifold fermions and higher-pseudospin quasiparticles within analogous frameworks~\cite{kikuchi2022quantum,ezawa2017chiral,lepori2018axial,nandy2019generalized,kikuchi2025band,balduini2024intrinsic,singh2023geometrical,mukherjee2025anomalous,wang2017nonequilibrium,ahmad2025longitudinal,ghosh2024linear,mandal2025chiral,haidar2025reflections,mandal2025signatures,mandal2025distinguishing,mandal2026longitudinal,yamada2026classical,gupta2026localization}. However, standard quasiclassical approaches often involve simplifying assumptions such as constant relaxation time approximations, omission of orbital magnetic moment (OMM) contributions, neglect of internode scattering processes, assuming the azimuthal symmetric setup of external electric $\mathbf{E}$ and magnetic fields $\mathbf{B}$, and disregard for global charge conservation constraints ~\cite{ahmad2025longitudinal,mandal2025chiral,mandal2025longitudinal}. These limitations can lead to significant deviations from experimentally observed behaviors, particularly in the experimentally relevant weak-field regime. It is therefore imperative to advance beyond these conventional assumptions to capture the full complexity of chiral anomaly-induced magnetotransport in chiral Fermionic systems. This necessity has been highlighted in recent works that incorporate more comprehensive theoretical treatments~\cite{knoll2020negative,sharma2020sign,ahmad2021longitudinal,ahmad2023longitudinal,ahmad2025chiral,sharma2023decoupling}. 

In this work, we present a comprehensive quasiclassical investigation of the chiral anomaly, planar Hall conductivity (PHC)~\cite{goldberg1954new}, and longitudinal magnetotransport in pseudospin-1 Fermionic systems, explicitly incorporating azimuthal symmetry-breaking factors such as non-collinear electric and magnetic fields, as well as tilting effects in the nonzero dispersive bands. By going beyond the conventional constant relaxation-time approximation and systematically accounting for both the orbital magnetic moment (OMM) contributions and global charge conservation constraints, we demonstrate that pseudospin-1 fermions exhibit a characteristic positive and quadratic-in-$B$ longitudinal magnetoconductivity (LMC) and PHC at low internode scattering strengths. Remarkably, as the internode scattering strength $\alpha$ exceeds a critical threshold $\alpha_c$, both LMC and PHC undergo a sign reversal, becoming negative. Notably, the critical value $\alpha_c$ at which this transition occurs is found to be lower for LMC than for PHC, highlighting the differential sensitivity of these transport coefficients to intervalley scattering processes. We study the angular dependence of PHC on magnetic field direction in the absence as well as the presence of tilt in the dispersion relation, and find that they are all different. While LMC shows a monotonic(non-monotonic) nature with respect to tilt $t_z (t_x)$, PHC does not follow the same trend and is seen to demonstrate non-monotonic dependence irrespective of tilt direction. As far as we know, PHC has never been studied in the pseudospin-1 Fermionic system, making this study unique and timely. As the real system possesses the dispersion, which may be tilted along a particular direction, this theoretical framework gains particular relevance in light of recent experimental studies probing chiral anomaly signatures in multifold Fermionic systems~\cite{balduini2024intrinsic,mukherjee2025anomalous}, as well as in anticipation of forthcoming experimental investigations on candidate materials crystallizing in crystallographic space groups $199$ ($I2_13$), $214$ ($I4_132$), and $220$ ($I–43d$)~\cite{bradlyn2016beyond}. These space groups correspond to the chiral tetrahedral ($23$) and chiral octahedral ($432$) point groups in crystal class terminology (see Table~\ref{tab:crystal_classes}), which are known to host multifold fermions characterized by nontrivial topological properties such as higher Chern numbers.
\begin{table}[b]
    \label{tab:crystal_classes}
    \begin{ruledtabular}
    \begin{tabular}{cccc}
    SG & H--M Symbol & PG & Sch. \\
    \hline
    199 & $I2_13$  & $23$  & $\mathrm{T}$ \\
    214 & $I4_132$ & $432$ & $\mathrm{O}$ \\
    220 & $I–43d$ & $432$ & $\mathrm{O_h}$ \\
    \end{tabular}
    \end{ruledtabular}
    \caption{Crystallographic space groups and point groups supporting multifold fermions that are relevant for experimental verification of the proposed theoretical predictions. SG denotes the space-group number, H-M Symbol the Hermann-Mauguin notation, PG the crystallographic point group, and Sch. the corresponding Schoenflies symbol.}
\end{table}
\section{Semimetals with 3-fold degeneracy}
\label{Semimetals with 3-fold degeneracy}
%\textit{Semimetals with 3-fold degeneracy:} 
The low-energy behavior of pseudospin-1 fermions in the presence of a directional tilt can be formulated within a field-theoretic framework. The corresponding Lagrangian density for a three-component spinor field $\Psi(\mathbf{x})$ is given by~\cite{peskin2018introduction,wilczek1999quantum}:
\begin{equation}
\mathcal{L} = \Psi^\dagger(\mathbf{x}) \left[ i\hbar \partial_t + i \hbar v_F \left( \mathbf{S} \cdot \nabla + (\mathbf{t} \cdot \nabla) P_d \right) \right] \Psi(\mathbf{x}),
\label{Lagrangian}
\end{equation}
where \( v_F \) is the material dependent velocity parameter, \( \mathbf{S} = (S_x, S_y, S_z) \) is the vector of spin-1 Pauli matrices matrices, and \( \mathbf{t} = t_x \hat{x} + t_z \hat{z} \) introduces tilt along specific directions in momentum space. The projection operator \( P_d \) isolates the dispersive subspace of the pseudospin-1 Hamiltonian and is defined as: $P_d = \sum_{\lambda = \pm 1} |\lambda\rangle \langle \lambda|$, where \( |\lambda\rangle \) are the eigenstates of \( \mathbf{S} \cdot \hat{\mathbf{k}} \) corresponding to eigenvalues \( \lambda = \pm 1 \). The projector onto the dispersive subspace of the pseudospin-1 Hilbert space is given by:
\begin{equation}
P_d = 
\begin{pmatrix}
1 & 0 & 0 \\
0 & 0 & 0 \\
0 & 0 & 1
\end{pmatrix},
\label{pmatrix}
\end{equation}
which excludes the flat band component associated with the eigenvalue $\lambda = 0$. To incorporate electromagnetic interactions, we perform minimal coupling $\partial_\mu \rightarrow D_\mu = \partial_\mu + i\frac{e}{\hbar} A_\mu$, where $A_\mu = (\phi, -\mathbf{A})$ is the electromagnetic gauge field. The gauge-invariant Lagrangian becomes:
\begin{equation}
\mathcal{L} = \Psi^\dagger(\mathbf{x}) \left[ i\hbar D_t + i \hbar v_F \left( \mathbf{S} \cdot \mathbf{D} + \mathbf{t} \cdot \mathbf{D} ~P_d \right) \right] \Psi(\mathbf{x}),
\label{GaugeLagrangian}
\end{equation}
which serves as the starting point for analyzing transport phenomena using semiclassical or Kubo formalism. In momentum space, the Hamiltonian associated with Eq.~\eqref{Lagrangian} takes the form
\begin{equation}
H(\mathbf{k}) = \hbar v_F \left( \mathbf{k} \cdot \mathbf{S} + \mathbf{t} \cdot \mathbf{k}~ P_d \right).
\label{FieldTheoryHamiltonian}
\end{equation}
Diagonalizing this Hamiltonian yields the energy eigenvalues:
\begin{equation}
\epsilon_{\mathbf{k}} = 0, \quad \pm \hbar v_F |\mathbf{k}|\left( 1 + \mathbf{t} \cdot \hat{\mathbf{k}} \right),
\label{DispersionRelation}
\end{equation}
where \( \hat{\mathbf{k}} = \mathbf{k}/|\mathbf{k}| \) is the unit vector in the direction of momentum. The energy spectrum comprises three bands, including a flat band at zero energy. Here, we consider that the flat band at zero energy remains unaffected by the tilt, while the two linearly dispersing bands acquire an asymmetric energy shift proportional to the component of the tilt along the momentum direction. The Bloch states corresponding to the two dispersive bands with nonzero energies are given by:
\begin{align}
|u^+\rangle &= \left[\cos^2\left(\frac{\theta}{2}\right) e^{-2i\phi}, ~ \frac{1}{\sqrt{2}}\sin(\theta) e^{-i\phi},  ~\sin^2\left(\frac{\theta}{2}\right)\right]^\mathrm{T}, \nonumber\\
|u^-\rangle &= \left[\sin^2\left(\frac{\theta}{2}\right) e^{-2i\phi}, ~ \frac{-1}{\sqrt{2}}\sin(\theta) e^{-i\phi},  ~\cos^2\left(\frac{\theta}{2}\right)\right]^\mathrm{T},
\label{Eq:wave_function}
\end{align}
where $\theta$ and $\phi$ denote the polar and azimuthal angles of the momentum vector, respectively, i.e., $\mathbf{k} = k(\sin \theta \cos \phi, \sin \theta \sin \phi, \cos \theta)$ and is measured from the nodal point. The total Berry flux associated with the two dispersive bands is quantized as $\nu = \mp 2$ (mod $2\pi$), indicating a monopole charge of $\pm 2$ at the nodal point. In contrast, the flat band is topologically trivial, contributing zero Berry curvature. 

Pseudospin-1 fermions may emerge as low-energy excitations in certain crystalline structures, such as body-centered cubic lattices with space groups 199, 214, and 220, where they appear at the $\mathbf{P}$ point of the Brillouin zone~\cite{bradlyn2016beyond}. Notably, the $\mathbf{P}$ point is not invariant under time-reversal symmetry, i.e., $\mathbf{P} \neq -\mathbf{P}$. According to the Nielsen–Ninomiya theorem~\cite{nielsen1981no,nielsen1983adler}, the net monopole charge in the Brillouin zone must vanish, requiring that sources and sinks of Berry curvature occur in compensating pairs. As a result, a minimal low-energy model must include contributions from opposite chiralities. Taking into account the possibility of a band tilt that can break time-reversal and Lorentz invariance and introduce spectral asymmetry, the effective Hamiltonian for a tilted pseudospin-1 semimetal can be expressed as \cite{yan2017topological, ahmad2024geometry,das2019linear,ahmad2025chiral,lv2015experimental}:
\begin{equation}
    H(\mathbf{k}) = \sum_{\chi = \pm 1}\hbar v_F \left(\chi \mathbf{k} \cdot \mathbf{S} + \mathbf{t}^\chi \cdot \mathbf{k}~ P_d \right),
    \label{Eq:Hamiltonian}
\end{equation}
where $\chi = \pm 1$ labels the chirality of the node, $\mathbf{t}^\chi$ is the tilt vector associated with each Weyl point, and $P_d$ is the projector onto the dispersive subspace of the pseudospin-1 Hilbert space (Please see Eq.~\ref{pmatrix}). This form ensures that only the dispersive bands acquire an energy shift due to the tilt, while the flat band remains unaffected. The Berry curvature associated with the conduction band of the Hamiltonian is given by: $ \boldsymbol{\Omega}^\chi_\mathbf{k} = -\chi \frac{\mathbf{k}}{k^3}$. The monopole-like distribution of $\boldsymbol{\Omega}^\chi_\mathbf{k}$ reflects the topological nature of the band crossing. Unlike a classical point particle, a Bloch wave packet in a periodic crystal possesses a finite spatial extent. As a result, it undergoes self-rotation about its center of mass, giving rise to an intrinsic orbital magnetic moment (OMM). This is expressed as~\cite{xiao2010berry,hagedorn1980semiclassical,chang1996berry}: $\mathbf{m}^{\chi}_\mathbf{k} = -\chi \frac{e v_F \mathbf{k}}{k^2}$. The orbital magnetic moment is aligned with the Berry curvature but exhibits a slower decay in momentum space. In the presence of an external magnetic field $\mathbf{B}$, the orbital magnetic moment couples to the field and modifies the band dispersion as: $\epsilon^\chi(\mathbf{k}) \rightarrow \epsilon(\mathbf{k}) - \mathbf{m}^\chi(\mathbf{k}) \cdot \mathbf{B}$. For the tilted pseudospin-1 semimetal, where the dispersion already includes an asymmetric tilt term, this coupling leads to a further anisotropic deformation of the Fermi surface.
Incorporating both the tilt and the orbital magnetic moment (OMM), the modified dispersion for the conduction band near a node of chirality $\chi$ in the presence of a magnetic field $\mathbf{B} = B (\cos\gamma\, \hat{x} + \sin\gamma\, \hat{z})$ is given by:
\begin{align}
\epsilon^\chi(\mathbf{k}) &= \hbar v_F k \left(1
+  t_x^\chi \sin\theta \cos\phi + t_z^\chi \cos\theta \right)\nonumber \\
&+ \chi \frac{e v_F B}{k} \left( \cos\gamma\, \sin\theta \cos\phi + \sin\gamma\, \cos\theta \right).
\end{align}
The tilt contribution appears as a linear energy shift, while the OMM correction is chirality-dependent and introduces inverse-momentum anisotropy. Solving $\epsilon^\chi(\mathbf{k}) = \epsilon_F$ for $k$ leads to the Fermi surface:
\begin{align}
k_F^{\chi}(\theta, \phi, \gamma) = \frac{\epsilon_F + \sqrt{\epsilon_F^2 - 4\eta\, t_{\text{fact}}\, \mathcal{G}_{m}^{\text{fact}}}}{2\, t_{\text{fact}}},
\end{align}
where, $\mathcal{G}_m = \cos(\theta)\sin(\gamma) + \cos(\gamma)\sin(\theta)\cos(\phi),
~~\mathcal{G}_{m}^{\text{fact}} = \chi e v_F B\, \mathcal{G}_m,~~t_{\text{fact}} = \hbar v_F \left[1 + t^{\chi}_{x} \sin(\theta)\cos(\phi) + t^{\chi}_{z} \cos(\theta)\right],$ and $\eta \in \{0, 1\}$ is a control parameter introduced to selectively toggle the effect of the orbital magnetic moment. The resulting Fermi surface becomes anisotropic due to both the tilt and the directional coupling of the OMM to the magnetic field. The velocity components corresponding to the conduction bands are evaluated as:
\begin{align}
v^{\chi}_{x}(\theta, \phi) &= t^{\chi}_{x} v_F + v_F \sin(\theta)\cos(\phi) \nonumber \\
&\quad + v^{\chi}(\theta, \phi)\left[\cos(\gamma)\left(1 - 2\sin^2(\theta)\cos^2(\phi)\right) \right. \nonumber \\
&\qquad \left. - 2\sin(\gamma)\sin(\theta)\cos(\theta)\cos(\phi)\right],\nonumber \\
v^{\chi}_{y}(\theta, \phi) &= v_F \sin(\theta)\sin(\phi) \nonumber \\
&\quad + v^{\chi}(\theta, \phi)\left[\cos(\gamma)\left(-2\sin^2(\theta)\sin(\phi)\cos(\phi)\right) \right. \notag \\
&\qquad \left. - 2\sin(\gamma)\sin(\theta)\cos(\theta)\sin(\phi)\right], \nonumber\\
v^{\chi}_{z}(\theta, \phi) &= t^{\chi}_{z} v_F + v_F \cos(\theta) \notag \\
&\quad + v^{\chi}(\theta, \phi)\left[\cos(\gamma)\left(-2\cos(\theta)\sin(\theta)\cos(\phi)\right) \right. \notag \\
&\qquad \left. + \sin(\gamma)\left(1 - 2\cos^2(\theta)\right)\right].
\end{align}
with, $ v^\chi(\theta, \phi) = \frac{\chi e \eta v_F B}{\hbar \left(k_F^\chi\right)^2}$.
%\begin{align}
%k_F^{\chi}(\theta, \phi, \gamma) = \frac{\epsilon_F + \sqrt{\epsilon_F^2 - 4\eta\, t_{\text{fact}}\, \mathcal{G}_{m}^{\text{fact}}}}{2\, t_{\text{fact}}},
%\end{align}

\section{Quasiclassical transport}
\label{Quasiclassical transport}
The semiclassical dynamics of quasiparticles in the presence of electric ($\mathbf{E}$) and magnetic ($\mathbf{B}$) fields are described by the following coupled equations~\cite{son2012berry,knoll2020negative,sundaram1999wave,son2012berry}:
\begin{align}
\dot{\mathbf{r}}^\chi &= \mathcal{D}^\chi_\mathbf{k} \left( \frac{e}{\hbar} \mathbf{E} \times \boldsymbol{\Omega}^\chi + \frac{e}{\hbar} \left( \mathbf{v}_\mathbf{k}^\chi \cdot \boldsymbol{\Omega}^\chi \right) \mathbf{B} + \mathbf{v}_\mathbf{k}^\chi \right), \nonumber \\
\dot{\mathbf{k}}^\chi &= -\frac{e}{\hbar} \mathcal{D}^\chi_\mathbf{k} \left( \mathbf{E} + \mathbf{v}_\mathbf{k}^\chi \times \mathbf{B} + \frac{e}{\hbar} (\mathbf{E} \cdot \mathbf{B}) \boldsymbol{\Omega}^\chi \right),
\label{Coupled_equation}
\end{align}
where $\mathbf{v}_\mathbf{k}^\chi = \frac{1}{\hbar} \frac{\partial \epsilon^\chi(\mathbf{k})}{\partial \mathbf{k}}$ denotes the band velocity, $\boldsymbol{\Omega}^\chi = -\chi \mathbf{k} / 2k^3$ is the Berry curvature, and $\mathcal{D}^\chi_\mathbf{k} = \left( 1 + \frac{e}{\hbar} \mathbf{B} \cdot \boldsymbol{\Omega}^\chi \right)^{-1}$ accounts for the modification of the phase-space volume due to Berry curvature effects~\cite{duval2006berry,xiao2010berry}. We would like to point out that we implicitly assume that the above semiclassical equations remain valid for higher pseudospin fermions as well, e.g., $s=1,3/2,...$. Although the effective kinetic theory can be derived from a Landau level basis for spin-1/2 fermions, such calculations have not yet been performed for pseudo-spin $>1/2$ fermions. However, we emphasize that the above quasiclassical equations of motion should remain independent of the pseudospin magnitude as they arise purely from band geometry. Although a microscopic Landau level derivation for higher spin-fermions is lacking, recent studies support the validity of this framework for capturing low-field magnetotransport in such systems~\cite{balduini2024intrinsic,singh2023geometrical,mukherjee2025anomalous,wang2017nonequilibrium,satow2025symmetry}. To analyze the transport properties of three-dimensional pseudospin-1 fermions under external electric and magnetic fields, we employ the quasiclassical Boltzmann formalism. The evolution of the non-equilibrium distribution function $f^{\chi}_{\mathbf{k}}$ is governed by~\cite{kim2014boltzmann,knoll2020negative,imran2018berry,zyuzin2017magnetotransport}:
\begin{align}
\frac{\partial f^{\chi}_{\mathbf{k}}}{\partial t} + \dot{\mathbf{r}}^{\chi}_{\mathbf{k}} \cdot \nabla_\mathbf{r} f^{\chi}_{\mathbf{k}} + \dot{\mathbf{k}}^{\chi} \cdot \nabla_\mathbf{k} f^{\chi}_{\mathbf{k}} = I_{\mathrm{coll}}[f^{\chi}_{\mathbf{k}}],
\label{MB_equation}
\end{align}
where $f^{\chi}_{\mathbf{k}} = f_0 + g^{\chi}_{\mathbf{k}}$, with $f_0$ denoting the equilibrium Fermi-Dirac distribution and $g^{\chi}_{\mathbf{k}}$ its perturbative deviation. To study the Planar Hall effect, we have fixed the electric field along $\hat{z}$-direction and magnetic field can be rotated in the $xz$-plane, i.e., $\mathbf{E} = E (0,0,1)$ and $\mathbf{B} = B (\cos \gamma,0,\sin \gamma)$. To linear order in the applied electric field, the deviation function takes the form~\cite{morimoto2016semiclassical,sodemann2015quantum,das2022nonlinear,mandal2022chiral,das2022nonlinear,gao2022suppression}: $g^{\chi}_{\mathbf{k}} = -e \left( \frac{\partial f_0}{\partial \epsilon} \right) \mathbf{\Lambda^{\chi}_{\mathbf{k}}} \cdot \mathbf{E}$, where $\mathbf{\Lambda^{\chi}_{\mathbf{k}}}$ is an unknown vector function determined from the Boltzmann equation. Since electric field is fixed along $\hat{z}$-direction, only $z$-component of $\mathbf{\Lambda^{\chi}_{\mathbf{k}}}$ is relevant. Therefor we can write~\cite{morimoto2016semiclassical,sodemann2015quantum,das2022nonlinear,mandal2022chiral}:
\begin{align}
g^{\chi}_{\mathbf{k}} = -e \left( \frac{\partial f_0}{\partial \epsilon} \right) \Lambda^{\chi}_{\mathbf{k},z} ~ E,
\label{Eq:g1}
\end{align}
Please note that for simplicity, here and onward, we will remove the lower index $z$ as it is not relevant. The collision integral $I_{\mathrm{coll}}[f^{\chi}_{\mathbf{k}}]$ accounts for impurity-induced scattering processes, comprising two distinct channels: (i) internode (intervalley) scattering, where $\chi \to \chi' \neq \chi$, and (ii) intranode (intravalley) scattering, where $\chi \to \chi' = \chi$. It is given by~\cite{mahan20089,bruus2004many,ziman1979principles,knoll2020negative,ahmad2024geometry,son2013chiral,kim2014boltzmann}:
\begin{align}
I_{\mathrm{coll}}[f^{\chi}_{\mathbf{k}}] = \sum_{\chi', \mathbf{k}'} W^{\chi \chi'}_{\mathbf{k k'}} \left( f^{\chi'}_{\mathbf{k'}} - f^{\chi}_{\mathbf{k}} \right),
\label{Collision_integral}
\end{align}
where the scattering rate $W^{\chi \chi'}_{\mathbf{k k'}}$ is calculated using Fermi's golden rule~\cite{ziman1979principles,bruus2004many,mahan20089,abers2004quantum}:
\begin{align}
W^{\chi \chi'}_{\mathbf{k k'}} = \frac{2\pi n}{\hbar \mathcal{V}} \left| \langle u^{\chi'}(\mathbf{k'}) | U^{\chi \chi'}_{\mathbf{k k'}} | u^{\chi}(\mathbf{k}) \rangle \right|^2 \delta \left( \epsilon^{\chi'}(\mathbf{k'}) - \epsilon^{\chi}(\mathbf{k}) \right),
\label{Fermi_golden_rule}
\end{align}
with $n$ denoting the impurity concentration, $\mathcal{V}$ the system volume, and $U^{\chi \chi'}_{\mathbf{k k'}}$ representing the impurity scattering potential. When evaluated at the Fermi level, $\epsilon^{\chi}_{\mathbf{k}} = \epsilon_F$, the scattering rate becomes purely angular in nature. For elastic impurities, we assume a momentum-independent scattering potential $U^{\chi \chi'}_{\mathbf{k k'}} = I_{3\times3} U^{\chi \chi'}$, where $U^{\chi \chi'}$ distinguishes between internode and intranode scattering amplitudes. The relative strength of chirality-flipping to chirality-conserving scattering processes is characterized by the ratio: ~$\alpha = U^{\chi \chi' \neq \chi}/U^{\chi \chi' = \chi}$. The overlap $\mathcal{T}^{\chi \chi'}({\theta,\theta',\phi,\phi'})=|\bra{u^{\chi'}(\mathbf{k'})}U^{\chi \chi'}_{\mathbf{k k'}}\ket{u^{\chi}(\mathbf{k})}|^2$ is generally a function of both the polar and azimuthal angles, making the scattering strongly anisotropic, calculated to be:
\begin{align}
&\mathcal{T}^{\chi \chi'}({\theta,\theta',\phi,\phi'}) \nonumber \\
&= |U^{\chi \chi'}|^2 \mathcal{V}^2 \bigg\{ 
\Big[ \frac{1}{4} + \frac{1}{2} \cos \theta \cos \theta' + \frac{1}{4} \cos^2 \theta \cos^2 \theta' \nonumber \\
&\quad + \frac{1}{2} \sin \theta \sin \theta' \cos(\phi - \phi') \nonumber \\ &\quad + \frac{1}{4} \sin^2 \theta \sin^2 \theta' \cos^2\phi \cos^2\phi'\nonumber \\
&\quad + \frac{1}{4} \sin^2 \theta \sin^2 \theta' \sin^2\phi \sin^2\phi'\nonumber \\
&\quad + \frac{1}{2} \cos \theta \cos \theta' \sin \theta \sin \theta' \cos(\phi - \phi') \Big] \delta_{\chi, \chi'} \nonumber \\
&\quad + \Big[ \frac{1}{4} - \frac{1}{2} \cos \theta \cos \theta' + \frac{1}{4} \cos^2 \theta \cos^2 \theta' \nonumber \\
&\quad - \frac{1}{2} \sin \theta \sin \theta' \cos(\phi - \phi')\nonumber \\ &\quad + \frac{1}{4} \sin^2 \theta \sin^2 \theta' \cos^2\phi \cos^2\phi'\nonumber \\
&\quad + \frac{1}{4} \sin^2 \theta \sin^2 \theta' \sin^2\phi \sin^2\phi'\nonumber \\ 
&\quad + \frac{1}{2} \cos \theta \cos \theta' \sin \theta \sin \theta' \cos(\phi - \phi') \Big] \delta_{\chi, -\chi'} \bigg\}.
\label{Overlap_of_spinor}
\end{align}
Using Eq's.~\ref{Coupled_equation}, \ref{Eq:g1} and \ref{Collision_integral}, Eq.~\ref{MB_equation} is written in the following form~\cite{bruus2004many,ziman1979principles,sharma2020sign,knoll2020negative,son2013chiral,sodemann2015quantum}:
\begin{align}
&\mathcal{D}^{\chi}_\mathbf{k}\left[{v^{\chi,z}_{\mathbf{k}}}+\frac{eB\sin \gamma}{\hbar}(\mathbf{v^{\chi}_k}\cdot\mathbf{\Omega}^{\chi}_k)\right]
 = \sum_{\chi' \mathbf{k}'}{\mathbf{W}^{\chi \chi'}_{\mathbf{k k'}}}{(\Lambda^{\chi'}_{\mathbf{k'}}-\Lambda^{\chi}_{\mathbf{k}})}.
 \label{Eq_boltz_E1}
 \end{align}
The primary objective of the present work is to isolate and investigate the contribution arising from the chiral-anomaly-induced transport channel. To this end, we neglect the conventional Lorentz-force contribution to the carrier dynamics. This approximation is justified within the weak-magnetic-field regime, where the relevant dimensionless expansion parameter is the cyclotron factor $\omega_c\tau \equiv \frac{eB}{m}\tau \ll 1$, with $\tau$ denoting the appropriate transport relaxation time. In this limit, the Lorentz-force term in the semiclassical equation of motion constitutes only a perturbative correction of order $\omega_c\tau$ relative to the electric-field-driven contribution and is therefore parametrically suppressed. In contrast, the chiral-anomaly-induced term proportional to $(\mathbf E \cdot \mathbf B) \boldsymbol{\Omega}^{\chi}_{\mathbf k}$ does not acquire such a suppression factor and can remain quantitatively significant even when $\omega_c\tau \ll 1$. For the parameter range relevant to Weyl metals, the condition $\omega_c\tau \ll 1$ is generally satisfied throughout the metallic regime where the semiclassical Boltzmann framework remains applicable. Consequently, retaining the anomaly-induced contribution while neglecting the Lorentz-force correction provides a controlled and physically well-motivated approximation. A lot of works can be found in this regime~\cite{varma2026chiral,li2021nonlinear,sharma2016nernst,nandy2021chiral,zeng2022chiral,sharma2023decoupling,ahmad2021longitudinal,das2019linear,gopalakrishnan2025chiral,varma2026strain}. Before further simplifying the above equation, we define the chiral scattering rate as follows:
\begin{align}
\frac{1}{\tau^{\chi}(\theta, \phi)}=\sum_{\chi'}\mathcal{V}\int\frac{d^3\mathbf{k'}}{(2\pi)^3}(\mathcal{D}^{\chi'}_{\mathbf{k}'})^{-1}\mathbf{W}^{\chi \chi'}_{\mathbf{k k'}}.
\label{Tau_invers}
\end{align}
Note the additional factor of $\mathcal{D^\chi_\mathbf{k}}$ due to the modification of density of states as a result of Berry curvature~\cite{duval2006berry}. The volume factor on the right-hand side of the above equation drops out when the $k$-space integration is performed. One might expect that for $\gamma = \pi/2$, where the electric and magnetic fields are both aligned along the $\hat{z}$-axis, azimuthal symmetry would be preserved \cite{ahmad2025longitudinal}. However, the presence of tilt in the Weyl cones explicitly breaks this azimuthal symmetry even when the electric and magnetic fields are parallel. Consequently, the integrals in the transport calculations, including the one defined in Eq.~\ref{Eq_boltz_E1}, as well as all subsequent angular integrations, must be performed over both the polar angle $\theta$ and the azimuthal angle $\phi$ whenever (i) the Weyl cones are tilted and/or (ii) $\gamma \neq \pi/2$. Furthermore, the radial integration remains simplified due to the delta function appearing in Eq.~\ref{Fermi_golden_rule}.
Eq.~\ref{Eq_boltz_E1} then transforms to:
\begin{align}
h^{\chi}(\theta, \phi) + \frac{\Lambda^{\chi}(\theta, \phi)}{\tau^{\chi}(\theta, \phi)}=\sum_{\chi'}\mathcal{V}\int\frac{d^3\mathbf{k}'}{(2\pi)^3} \mathcal{D}^{\chi'}_{\mathbf{k}'}\mathbf{W}^{\chi \chi'}_{\mathbf{k k'}}\Lambda^{\chi'}(\theta',\phi').
\label{MB_in_term_Wkk'}
\end{align}
\begin{figure*}
    \centering
    \includegraphics[width=1.98\columnwidth]{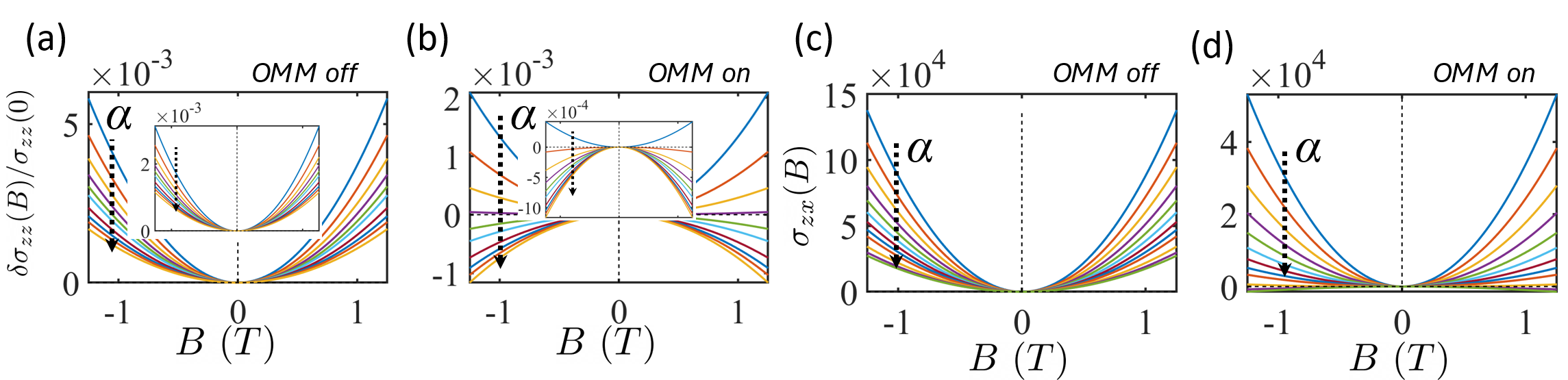}
    \caption{We investigate the LMC and PHC in a pseudospin-1 semimetal, both in the presence and absence of the OMM, for two representative values of the field orientation angle: $\gamma = \pi/2$ and $\gamma = \pi/4$. Our analysis reveals that an increase in the relative strength of internode (intervalley) scattering, quantified by the parameter $\alpha$, leads to a sign reversal in both LMC and PHC, from positive to negative. Furthermore, in configurations where the electric and magnetic fields are non-collinear, the critical scattering strength $\alpha_c$ required to induce this reversal decreases progressively as $\gamma$ deviates from the orthogonal limit ($\gamma = \pi/2$). This behavior underscores the sensitivity of magnetotransport characteristics to the interplay between field orientation and intervalley scattering in topological semimetals. (a-b) are for parallel electric and magnetic fields $\gamma=\pi/2$, while inset represents for $\gamma=\pi/4$. PHC at $\gamma=\pi/4$ is plotted in (c-d).}
    \label{fig:LMC_and_PHC_spin1.png}
\end{figure*}
Here, $h^{\chi}(\theta, \phi)=\mathcal{D}^{\chi}_{\mathbf{k}}[v^{\chi}_{z,\mathbf{k}}+eB \sin \gamma(\mathbf{\Omega}^{\chi}_{k}\cdot \mathbf{v}^{\chi}_{\mathbf{k}})/
\hbar]$, evaluated on the Fermi surface. Changing integration over momentum $k$ into energy integration using modified dispersion relation present in Eq.~\ref{Tau_invers} and Eq.~\ref{MB_in_term_Wkk'}, simplified to integration over $\theta'$ and $\phi'$:
\begin{align}
\frac{1}{\tau^{\chi}(\theta, \phi)} = \sum_{\chi'} \Pi^{\chi\chi'}\iint\frac{(k')^3\sin{\theta' d\theta' d\phi'}}{|\mathbf{v}^{\chi'}_{k'}\cdot{\mathbf{k'}^{\chi'}}|} \mathcal{T}^{\chi\chi'} (\mathcal{D}^{\chi'}_{\mathbf{k'}})^{-1}.
\label{Tau_inv_int_theta}
\end{align}
\begin{align}
h^{\chi}(\theta, \phi) &+ \frac{\Lambda^{\chi}(\theta, \phi)}{\tau^{\chi}(\theta, \phi)}\nonumber\\&=\sum_{\chi'} \Pi^{\chi\chi'}\iint d\theta'd\phi' ~f^{\chi'}(\theta',\phi') \times\nonumber\\&\quad\quad~\mathcal{T}^{\chi\chi'}(\theta,\theta',\phi,\phi')\Lambda^{\chi'}(\theta',\phi')/\tau^{\chi'}(\theta',\phi'),
\label{Eq_h_tau_boltz}
\end{align}
 where, $\Pi^{\chi \chi'} = n/4\pi^2 \hbar^2$ ($N=n\mathcal{V}$ being the number of impurities), $f^{\chi} (\theta, \phi)=\frac{(k)^3}{|\mathbf{v}^\chi_{\mathbf{k}}\cdot \mathbf{k}^{\chi}|} \sin\theta (\mathcal{D}^\eta_{\mathbf{k}})^{-1} \tau^\chi(\theta, \phi)$ and $\mathcal{T}^{\chi\chi'}$ is defined in Eq.~\ref{Overlap_of_spinor}. With the ansatz $\Lambda^{\chi}(\theta, \phi)= [\lambda^{\chi} - h^{\chi}_{\mu} 
+ a^{\chi}_{1} \cos(\theta) + a^{\chi}_{2} \cos^{2}(\theta) 
+ b^{\chi}_{1} \sin(\theta) \cos(\phi) 
+ b^{\chi}_{2} \sin(\theta) \sin(\phi) + c^{\chi}_{1} \cos(\theta) \sin(\theta) \cos(\phi) 
+ c^{\chi}_{2} \cos(\theta) \sin(\theta) \sin(\phi) + d^{\chi}_{1} \sin^{2}(\theta) \cos^{2}(\phi) 
+ d^{\chi}_{2} \sin^{2}(\theta) \sin^{2}(\phi) 
+ d^{\chi}_{3} \sin^{2}(\theta) \cos(\phi) \sin(\phi)]\tau^{\chi}(\theta, \phi)$, Eq.~\ref{Eq_h_tau_boltz} is expressed as:
 \begin{align}
&\lambda^{\chi}  + a^{\chi}_{1} \cos(\theta) + a^{\chi}_{2} \cos^{2}(\theta) 
+ b^{\chi}_{1} \sin(\theta) \cos(\phi) \nonumber\\
&+ b^{\chi}_{2} \sin(\theta) \sin(\phi) + c^{\chi}_{1} \cos(\theta) \sin(\theta) \cos(\phi) \nonumber\\
&+ c^{\chi}_{2} \cos(\theta) \sin(\theta) \sin(\phi) + d^{\chi}_{1} \sin^{2}(\theta) \cos^{2}(\phi) \nonumber\\
&+ d^{\chi}_{2} \sin^{2}(\theta) \sin^{2}(\phi) 
+ d^{\chi}_{3} \sin^{2}(\theta) \cos(\phi) \sin(\phi)\nonumber\\
&=\sum_{\chi'}\Pi^{\chi\chi'}\int f^{\chi'}(\theta')d\theta' ~\mathcal{T}^{\chi\chi'} ({\theta,\theta',\phi,\phi'}) \nonumber\\\times&[\lambda^{\chi} - h^{\chi}_{\mu} 
+ a^{\chi}_{1} \cos(\theta') + a^{\chi}_{2} \cos^{2}(\theta') 
+ b^{\chi}_{1} \sin(\theta') \cos(\phi') \nonumber\\
&+ b^{\chi}_{2} \sin(\theta') \sin(\phi') 
 + c^{\chi}_{1} \cos(\theta') \sin(\theta') \cos(\phi') \nonumber \\
&+ c^{\chi}_{2} \cos(\theta') \sin(\theta') \sin(\phi') 
+ d^{\chi}_{1} \sin^{2}(\theta') \cos^{2}(\phi') \nonumber \\
&+ d^{\chi}_{2} \sin^{2}(\theta') \sin^{2}(\phi') 
+ d^{\chi}_{3} \sin^{2}(\theta') \cos(\phi') \sin(\phi')].
\label{Boltzman_final}
\end{align}
When explicitly written, this equation consists of 20 simultaneous equations that need to be solved for 20 variables, and particle number conservation serves as an additional constraint, is given by~\cite{sharma2023decoupling}:
\begin{align}
\sum\limits_{\chi}\sum\limits_{\mathbf{k}} g^\chi_\mathbf{k} = 0 \label{Eq_sumgk}
\end{align}
The current is calculated using the expression :
\begin{align}
    \mathbf{J}=-e\sum_{\chi,\mathbf{k}} f^{\chi}_{\mathbf{k}} \dot{\mathbf{r}}^{\chi}.
    \label{Eq:J_formula}
\end{align}
Due to the high dimensionality and the intricate functional dependence on angular variables, all the coupled equations and the integrals over the polar ($\theta$) and azimuthal ($\phi$) angles have been evaluated using numerical methods. Specifically, discretization schemes and symbolic calculation techniques were employed to accurately capture the angular dependence and solve the system self-consistently.
\section{Results and Discussion}
\label{Results and Discussion}
We investigate the chiral anomaly-induced transport characteristics of pseudospin-1 fermions under broken azimuthal symmetry, implemented through two distinct mechanisms: (i) the application of a rotating magnetic field confined to the $xz$-plane, and (ii) the incorporation of an anisotropic tilt in the energy dispersion relation. This configuration enables a systematic analysis of the chiral-anomaly-induced influence on both LMC and PHC. The next sections present a detailed analysis of the corresponding results for each scenario.

\subsection{LMC and PHC in pseudospin-1 fermions for collinear electric and magnetic field}
We initiate our analysis by examining the LMC characteristics of a pseudospin-1 semimetal subjected to collinear electric and magnetic fields ($\mathbf{E} \parallel \mathbf{B}$) within the semiclassical Boltzmann transport framework. To ensure consistency with prior findings while preserving the system's inherent azimuthal symmetry~\cite{ahmad2025longitudinal}, we specifically consider this field configuration. The alignment $\mathbf{E} \parallel \mathbf{B}$ confines charge carrier dynamics predominantly along the $z$-axis, thereby simplifying the problem analytically by suppressing transverse current contributions. Notably, in this configuration, the Lorentz force $\mathbf{F} \propto \mathbf{v} \times \mathbf{B}$ identically vanishes, eliminating conventional Hall-type responses. The tensorial form of the magnetoconductivity is captured via a perturbative expansion in the magnetic field strength $B$ as follows~\cite{ahmad2023longitudinal,ahmad2021longitudinal,sharma2020sign}:
\begin{align}
\sigma_{ij}(B) = \sigma^{(2)}_{ij} \, B^2 + \sigma^{(1)}_{ij} \, B + \sigma^{(0)}_{ij},
\label{Eq: conductivity tensor ij}
\end{align}
where $\sigma^{(n)}_{ij}$ denote the field-independent coefficients associated with the $n$-th order magnetic field contributions to the conductivity tensor components, which may be a function of different parameters, like tilt, intervalley scattering, as well as the magnetic field's angle $\gamma$. This functional dependence encapsulates both intrinsic and extrinsic magnetotransport phenomena, including possible contributions from Berry curvature effects and scattering mechanisms. In Fig.~\ref{fig:LMC_and_PHC_spin1.png}(a) and (b), we present the LMC, normalized by its zero-field value, as a function of the applied magnetic field strength, i.e., $\delta\sigma_{zz} =(\sigma_{zz}(B) - \sigma_{zz}(B=0))/\sigma_{zz}(B=0)$. The LMC exhibits a quadratic dependence on the magnetic field. For $\alpha < \alpha_c$, where $\alpha$ denotes the relative internode scattering strength and $\alpha_c$ is its critical threshold value, the LMC remains positive. However, upon surpassing this critical value ($\alpha > \alpha_c$), a sign reversal in the LMC is observed, indicating a transition from positive to negative conductivity. Notably, in the absence of the orbital magnetic moment contribution, the LMC remains strictly positive across all values of $\alpha$. In this case, increasing $\alpha$ merely attenuates the magnitude of the conductivity without inducing any sign change. This is in accordance with our previous results ~\cite{ahmad2025longitudinal}. In this geometry of the fields, the PHC is zero as expected. To push the existing knowledge, we will now break the azimuthal symmetry by taking $\gamma\ne \pi/2$.
\begin{figure}
    \centering
    \includegraphics[width=1\columnwidth]{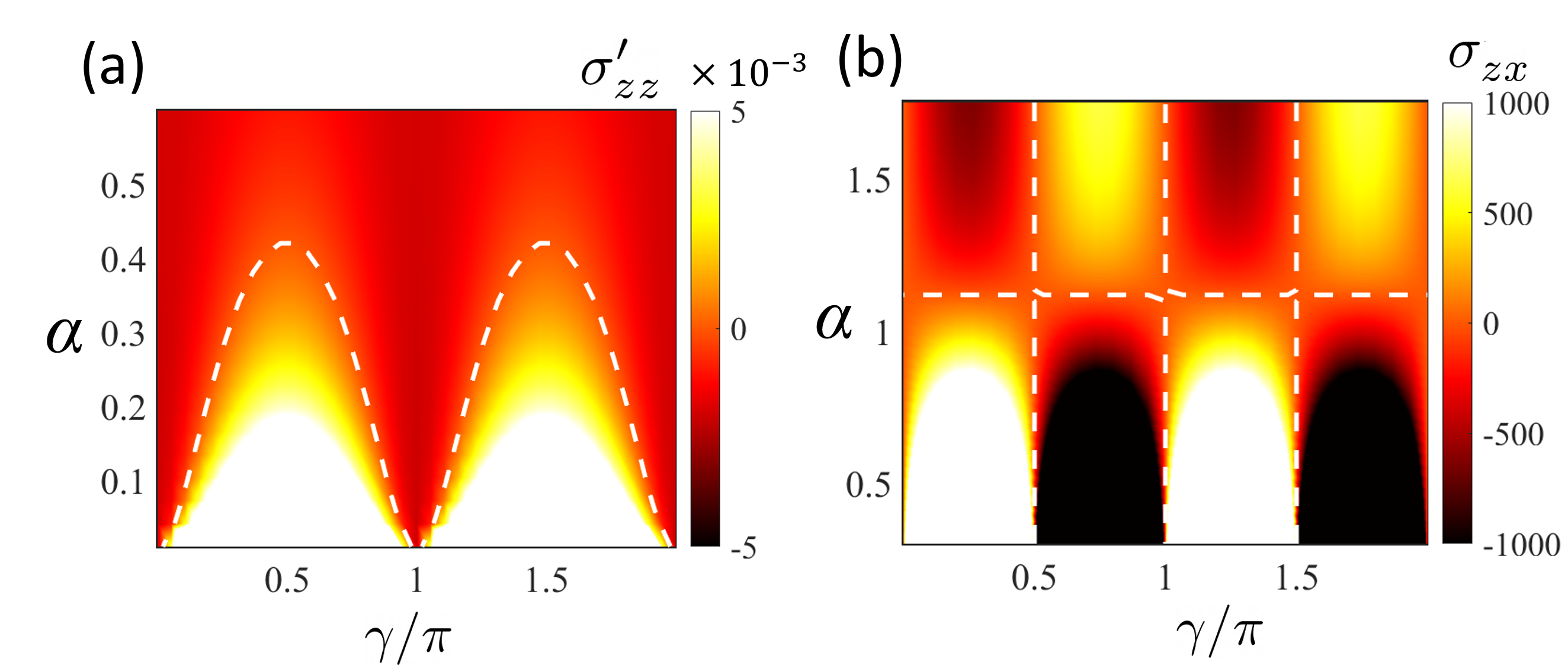}
    \caption{Phase plot of LMC and PHC for pseudospin-1 fermions with respect to intervalley scattering strength $\alpha$ and magnetic field angle $\gamma$ measured from increasing x-direction. While LMC has its maxima at $\pi/2$, PHC vanishes for this parallel field configuration. White dotted curves separate the region having opposing signs of conductivities, confirming the dependence $\alpha_c(\gamma)$: $\alpha_c \sim \sin^2 \gamma$ in LMC. PHC vanishes for both the field orientations: $\gamma=0$ and $\gamma=\pi/2$ as plotted in (b), shows sign reversal above a fixed value of $\alpha_c^{PHC}>\alpha_c^{LMC}$. LMC is normalized by zero field value, i.e., $\sigma_{zz}^{'} = (\sigma_{zz} (B=0.50 ~T) - \sigma_{zz} (B=0 ~T))/\sigma_{zz} (B=0 ~T) $, while PHC is not normalized.}
    \label{fig:LMC_and_PHC_spin1_phase_plot_tilt0.png}
\end{figure}
\subsection{LMC and PHC in pseudospin-1 fermions for non-collinear electric and magnetic field}
We now explicitly break the azimuthal symmetry of the system by tilting the magnetic field away from the direction of the electric field. The insets of Fig.~\ref{fig:LMC_and_PHC_spin1.png}(a) and (b) display the LMC for the case $\gamma = \pi/4$. It is observed that the qualitative behavior of the LMC remains analogous to the $\gamma = \pi/2$ configuration; however, the critical relative internode scattering strength $\alpha_c$ at which the sign reversal occurs is quantitatively reduced, thereby confirming the functional dependence $\alpha_c = \alpha_c(\gamma)$. To get more insight, we study the LMC as function of $\alpha$ and $\gamma$ in Fig.~\ref{fig:LMC_and_PHC_spin1_phase_plot_tilt0.png}. In the limit $\alpha \to 0^+$, the LMC consistently exhibits a positive sign irrespective of the variation in the angle $\gamma$, showing no signature of sign reversal. The inclusion of intervalley scattering mechanisms induces a qualitative modification in the LMC behavior, manifesting as a sign reversal within the parameter range $\gamma \in (0, \pi/2)$. Furthermore, we observe that the critical intervalley scattering value $\alpha_c$ attains its minimum at $\gamma = 0$, increases monotonically with $\gamma$, and reaches its maximal value at $\gamma = \pi/2$. We note that $\alpha_c \sim \sin^2 \gamma$ and it hasn't been reported before.
\begin{figure*}
    \centering
    \includegraphics[width=1.98\columnwidth]{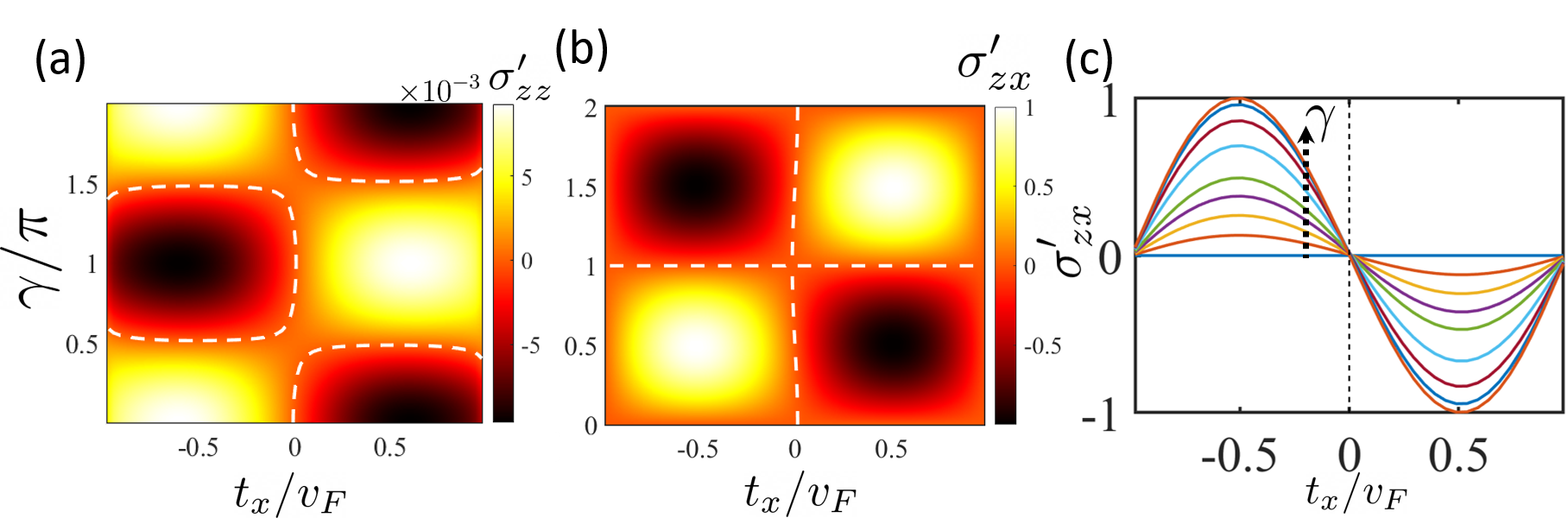}
    \caption{(a-b) The LMC and PHC are analyzed as functions of the tilt along the $x$-axis and the orientation angle $\gamma$ of the applied magnetic field relative to the $\hat{x}$-axis, for a constant magnetic field strength of $B = 0.5~\text{T}$ and $\alpha = 0.20$. In this analysis, the cone tilt directions are assumed to be oppositely oriented. (c) PHC is plotted as a function of tilt $t_x$. PHC shows a non-monotonic behavior and is an odd function of tilt, attaining its extrema at $|t_x|=0.50$. Along the direction of the arrow, the $\gamma$ is varied from 0 to $\pi/2$. PHC is normalized to its maximum value and LMC as in Fig.~\ref{fig:LMC_and_PHC_spin1_phase_plot_tilt0.png}(a).}
    \label{fig:LMC and PHC vs tx and GM.png}
\end{figure*}
Furthermore, in Fig.~\ref{fig:LMC_and_PHC_spin1.png}(c) and (d), we present the PHC as a function of the magnetic field strength $B$ for various values of $\alpha$. Similar to the LMC, the PHC also undergoes a sign reversal from positive to negative as $\alpha$ increases. Importantly, the presence of the OMM is essential for this sign change to manifest. We also note that the critical scattering strength $\alpha_c$ required for the sign reversal in PHC is comparatively larger than that for the LMC. In contrast to the LMC, the critical parameter $\alpha_c$ for PHC exhibits only a weak dependence on the parameter $\gamma$, as illustrated in Fig.~\ref{fig:LMC_and_PHC_spin1_phase_plot_tilt0.png}(b). Notably, the PHC remains strictly positive within the entire range $\gamma \in (0, \pi/2)$, undergoing a sign reversal exclusively under conditions of sufficiently strong intervalley scattering strength($\alpha_c\simeq1.20$), indicative of a dominant inter-valley relaxation mechanism.
\begin{figure*}
    \centering
    \includegraphics[width=1.98\columnwidth]{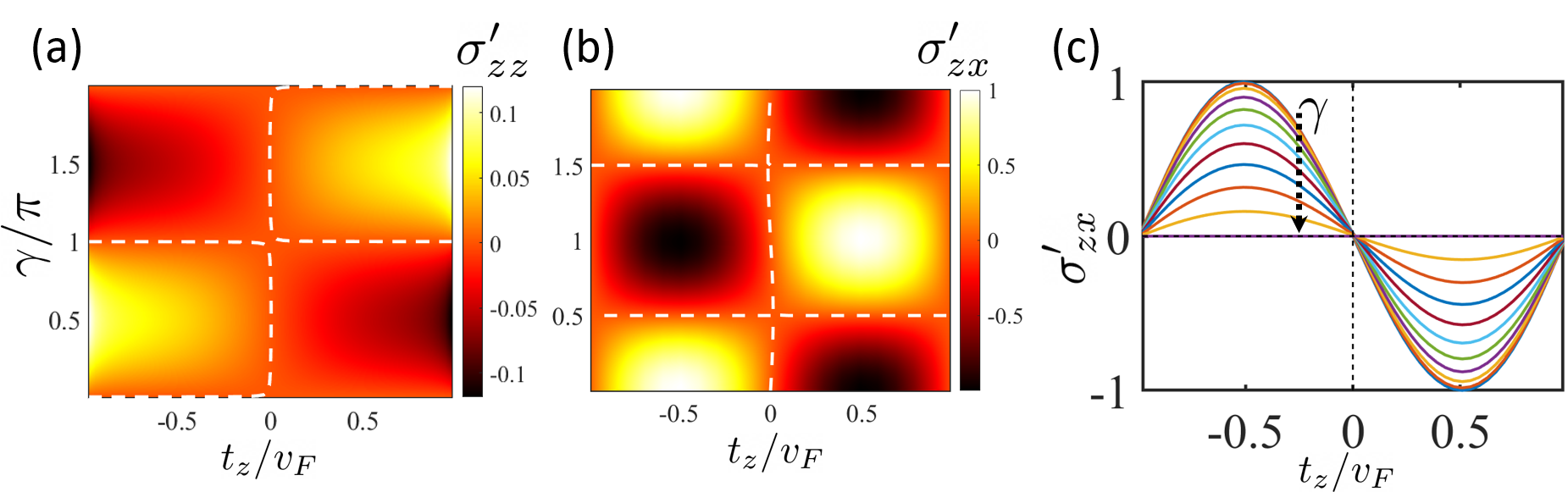}
    \caption{ (a-b) The LMC and PHC are analyzed as functions of the tilt along the $z$-axis and the orientation angle $\gamma$ of the applied magnetic field relative to the $\hat{z}$-axis, for a constant magnetic field strength of $B = 0.5~\text{T}$. In this analysis, the tilt directions of the cones are considered to be oppositely oriented. (c) PHC as a function of tilt $t_z$ at different values of $\gamma$.}
    \label{fig:LMC and PHC vs tz and GM.png}
\end{figure*}
\subsection{LMC and PHC in pseudospin-1 fermions for tilted Weyl cone}
We consider a systematic investigation of two distinct tilt configurations in tilted pseudospin-1 Weyl semimetals (WSMs): (i) identical tilt orientation, characterized by $t^{\chi}_{\hat{i}} = t^{\chi'}_{\hat{i}}$, and (ii) opposite tilt orientation, given by $t^{\chi}_{\hat{i}} = -t^{\chi'}_{\hat{i}}$, where $i \in \{x, z\}$ denotes the spatial directions under consideration. In the case of oppositely tilted Weyl cones, the linear term in $B$, in the conductivity tensor, $\sigma_{zz}$, persists, resulting in a titlted parabolic profile of the LMC as a function magnetic field $B$. In Fig.~\ref{fig:LMC and PHC vs tx and GM.png}(a), we present the normalized $\sigma_{zz}(\gamma, t_x, B = 0.5~\mathrm{T})$ as a function of the tilt component $t_x$ and the angle $\gamma$, specifically for the opposite tilt orientation. Our analysis reveals that, for a fixed $\gamma\in(0,\pi)$, increasing the tilt parameter induces a non-monotonic behavior in the magnitude of $\sigma_{zz}$. This exhibits anisotropic characteristics: for tilt along the $x$-axis, the $\gamma$-dependence is distinct in comparison to tilt along the $z$-axis, a more pronounced and monotonic dependence emerges, as depicted in Fig.~\ref{fig:LMC and PHC vs tz and GM.png}(a). This anisotropic behavior underscores the directional dependence of $\sigma_{zz}$ on the relative alignment between the tilt vector $\mathbf{t}$ and the external magnetic field $\mathbf{B}$.\\

\begin{figure}
    \centering
    \includegraphics[width=1\columnwidth]{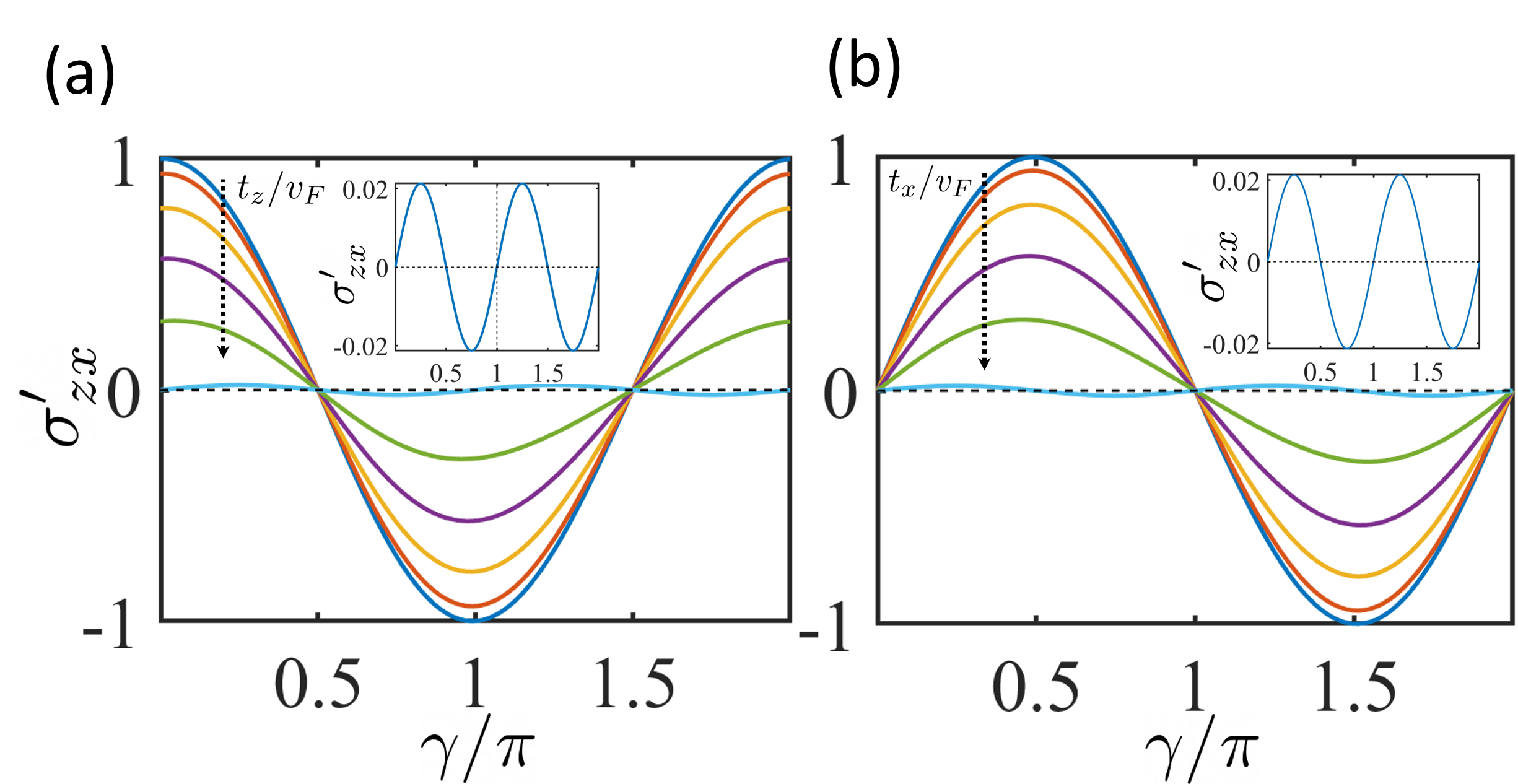}
    \caption{(a) The angular dependence of the PHC is investigated for varying values of the tilt parameter $t_z$. The inset corresponds to the untilted case ($t_z = 0$), exhibiting the conventional $\sin 2\gamma$ dependence characteristic of systems with preserved azimuthal symmetry. Introducing a finite tilt modifies this angular behavior, progressively transitioning towards a $\cos \gamma$ dependence as $t_z$ increases in magnitude. Specifically, along the indicated direction in the plot, the ratio $t_z / v_F$ is systematically varied from $-0.5$ to $0$, where the tilt orientation is considered to be oppositely aligned for the two cones, emphasizing the role of tilt anisotropy in shaping the PHC response. (b) For tilt along the $x$-direction. Unlike tilt along $z$-direction, there is a change of angular dependence, i.e., in the presence of tilt, the PHC now varies as $\sin\gamma$.}
    \label{fig:PHC vs gamma tz vary tz vary.png}
\end{figure}

Furthermore, we investigate the PHC in tilted pseudospin-1 Weyl semimetals as a function of the angle $\gamma$, with the corresponding results illustrated in Figs.~\ref{fig:LMC and PHC vs tx and GM.png}(b) and (c). Notably, PHC vanishes when the tilt is absent, and the electric and magnetic fields are orthogonal. As a function of $t_x$, the PHC exhibits non-monotonic behavior and is an odd function of the tilt parameter. The maximal PHC is observed for $\gamma \to\pi/2$ and $t_x \to 0.5$, which is explicitly plotted in Fig.~\ref{fig:LMC and PHC vs tx and GM.png}(c). Additionally, we analyze the angular dependence of PHC on the orientation of the magnetic field $\mathbf{B}$, which is experimentally accessible, as shown in Fig.~\ref{fig:PHC vs gamma tz vary tz vary.png}(b). Our findings establish that for tilted pseudospin-1 fermionic quasiparticle excitations, the off-diagonal conductivity component follows $\sigma_{zx} \propto \sin\gamma$. It is important to highlight that in the absence of tilt, the angular dependence transitions to $\sigma_{zx} \propto \sin 2\gamma$, as depicted in the insets of Fig.~\ref{fig:PHC vs gamma tz vary tz vary.png}. These results provide critical insights into the anisotropic magnetotransport phenomena in tilted pseudospin-1 Weyl systems.\\

After analyzing the PHC for the case where the Weyl cones are tilted along the $x$-axis, for the sake of completeness, we now extend our investigation to the scenario in which the tilt is oriented along the $z$-axis. While it may be intuitively expected that the conductivity response remains qualitatively similar across different tilt directions, our results reveal subtle yet significant distinctions. In Fig.~\ref{fig:LMC and PHC vs tz and GM.png}(b), we present the computed PHC for Weyl cones tilted along the $z$-direction. Contrary to the $x$-tilted case, the angular dependence of PHC in this configuration follows $\sigma_{zx} \propto \cos\gamma$. Furthermore, the PHC exhibits analogous non-monotonic behavior as a function of the tilt parameter $t_z$, consistent with the $x$-tilted scenario. This non-monotonicity is explicitly visualized in Fig.~\ref{fig:LMC and PHC vs tz and GM.png}(c) as well as in Fig.~\ref{fig:PHC vs gamma tz vary tz vary.png}(a). These observations underscore the anisotropic and tilt-direction-dependent nature of planar Hall responses in tilted pseudospin-1 Weyl semimetals. The dependence on $\cos\gamma$ rather than $\sin\gamma$ for $z$-tilted systems highlights a distinct symmetry behavior governed by the relative orientation of the tilt vector and the applied external magnetic field.
\section{Outlook and Conclusion}
\label{Outlook and Conclusion}
In this work, we have conducted a comprehensive theoretical analysis of LMC and PHC in pseudospin-1 WSMs, focusing on both untilted and tilted Weyl cone configurations under semiclassical Boltzmann transport theory. Our study systematically accounts for the effects of collinear and non-collinear electric and magnetic field orientations, intervalley scattering strength, and tilt-induced anisotropy. For untilted pseudospin-1 WSMs under collinear electric and magnetic fields ($\mathbf{E} \parallel \mathbf{B}$), we observe that LMC exhibits a positive quadratic dependence on the magnetic field in the presence of small intervalley scattering. The increment of intervalley scattering introduces a critical scattering strength $\alpha_c$, beyond which a sign reversal from positive to negative LMC occurs. This critical value is maximized at $\gamma = \pi/2$, highlighting its angular dependence $\alpha_c \sim \sin^2 \gamma$. As expected from symmetry considerations, PHC vanishes identically in this collinear configuration. When the azimuthal symmetry is explicitly broken by introducing a finite angle between $\mathbf{E}$ and $\mathbf{B}$, we observe that the qualitative behavior of LMC remains consistent; however, $\alpha_c$ decreases as $\gamma$ reduces from $\pi/2$, reaching its minimum for $\gamma = 0$. Additionally, PHC becomes finite in this geometry ($0<\gamma<\pi/2$) and similarly exhibits a sign reversal as a function of $\alpha$. Notably, the critical value $\alpha_c$ for PHC is larger than for LMC ($\alpha_c^{LMC}<\alpha_c^{PHC}$). Extending our analysis to tilted pseudospin-1 Weyl semimetals, we have demonstrated that the introduction of tilt leads to anisotropic modifications in both LMC and PHC. For oppositely tilted Weyl cones, the linear term in $\sigma_{ij}$ persists, producing a tilted parabolic conductivity profile with respect to the external magnetic field. Our results reveal distinct symmetry behaviors in PHC depending on the tilt orientation. Specifically, for tilt along the $x$-axis, PHC follows $\sigma_{zx} \propto \sin \gamma$, whereas for tilt along the $z$-axis, it obeys $\sigma_{zx} \propto \cos \gamma$. In both cases, PHC exhibits non-monotonic dependence on the tilt parameter and is maximized for specific tilt values and field orientations. Importantly, in the absence of tilt, the angular dependence of PHC reduces to $\sigma_{zx} \propto \sin 2\gamma$, confirming the intrinsic tilt-induced anisotropy. Overall, our findings provide detailed insights into the interplay between tilt, intervalley scattering, and magnetic field orientation in governing magnetotransport properties of pseudospin-1 Weyl semimetals. These results not only extend the fundamental understanding of higher pseudospin systems but also offer experimentally relevant predictions for tuning LMC and PHC via external field configurations and material-specific tilt engineering.

\section{Acknowledgments}
I acknowledge financial support from ANRF-SERB Core Research Grant CRG/2023/005628. I am very grateful to Gargee Sharma, Amit Agarwal, Sunit Das, Gautham Varma K, Hridis Kumar Pal, Shantanu Mukherjee, Mohd. Hashim Raza, Shubhanshu Karoliya, Pankaj Bhalla, Snehasish Nandy, and Tanay Nag for insightful discussions and valuable scientific input.
\bibliography{biblio.bib}
\end{document}